\documentclass[reprint,amssymb,pra,longbibliography]{revtex4-2}
\usepackage[utf8]{inputenc}
\usepackage[T1]{fontenc}
\usepackage{graphicx}
\usepackage{amsmath}
\usepackage{amssymb}
\usepackage{mathrsfs}
\usepackage{braket}
\usepackage{geometry}
\usepackage{here}
\usepackage{lmodern}
\usepackage{array}
\usepackage{natbib}
\usepackage{url}
\usepackage{textcase}
\usepackage{bm}
\usepackage{natbib}
\usepackage{color}
\begin{document}
\title{Complete quantum coherent control of ultracold molecular collisions}

\author{Adrien Devolder$^{1}$, {Paul Brumer$^{1}$, and Timur V. Tscherbul$^{2}$}}

\affiliation{$^{1}$Chemical Physics Theory Group, Department of Chemistry, and Center for Quantum Information and Quantum Control, University of Toronto, Toronto, Ontario, M5S 3H6, Canada\\
$^{2}$Department of Physics, University of Nevada, Reno, NV, 89557, USA}

\begin{abstract}
We show that quantum interference-based coherent control is a highly efficient tool for tuning ultracold molecular collision dynamics, and is free from the limitations of commonly used methods that rely on external electromagnetic fields. By varying {the relative populations and} phases of an initial coherent superpositions of degenerate molecular states, we demonstrate complete coherent control over integral scattering cross sections in the ultracold $s$-wave regime of both the initial and final collision channels. The proposed control  methodology is applied to  ultracold O$_2$~+~O$_2$ collisions, showing extensive control over $s$-wave spin-exchange cross sections and product branching ratios over many orders of magnitude.

\end{abstract}

\date{\today}
	\maketitle
{\it Introduction.} Recent advances in experimental techniques for cooling and trapping neutral atoms and polar molecules \cite{Bohn:17,Liu:18,Norcia:18,Cooper:18} have reignited interest in novel approaches to controlling atomic and molecular collisions and chemical reactivity at  ultralow temperatures. Such approaches are central to  using ultracold atoms and molecules in optical lattices as a platform for quantum information processing and quantum simulation \cite{Bloch:08,Gross:17,Carr:09,Bohn:17} and to studying exotic regimes of ultracold controlled chemistry \cite{Krems:08,Balakrishnan:16,Bohn:17}.
Several experimental groups have prepared ultracold molecules in a single quantum state, \cite{Bohn:17} enabling the use of quantum statistics \cite{Ni:10}, external electric fields, scattering resonances \cite{Vogels:15,Yang:19}, and steric effects \cite{Perreault:17,Jambrina:19,Morita:20}  to control molecular collisions and chemical reactivity at ultralow temperatures. 


The vast majority of control scenarios developed thus far  for ultracold atomic and molecular collisions are based on a combination of  static (dc) and time-varying (ac) external electromagnetic fields.  Examples include magnetic and optical Feshbach resonances \cite{Chin2010,Stwalley1976,Tiesinga1993,Fedichev1996,Devolder2019,Tscherbul2010},  electric field-induced resonances  \cite{Avdeenkov:03,Tscherbul:15}, microwave dressing \cite{Sanchez-Villicana1995,Avdeenkov2006,Alyabyshev2009,Quemener2016,Karman2018,Lassabliere2018,Xie2020}, parity breaking in superimposed electric and magnetic fields \cite{Abrahamsson2007,Tscherbul2006}, and  low-dimensional confinement   \cite{Li2008,Li2009,Miranda:11}.
Despite the success of these control methods, they suffer from a number of serious limitations.
First, dc fields control cannot be applied to control molecular systems that lack magnetic (or electric)
dipole moments, such as H$_2$. Such systems are often of great chemical and astrochemical interest, and have been  studied with unprecedented theoretical accuracy, such as the  archetypal chemical reaction F~+~H$_2$ $\to$ HF~+~H  \cite{Ren:08,Tizniti:14,DeFazio:20}.
 Second, the extent of control is limited by the magnitude of
molecular Stark and Zeeman shifts induced by practical laboratory dc fields, restricting the
scope of controllable molecules to those with large electric and magnetic dipole moments.
Finally, the presence of external field-induced perturbations can be counterproductive in high-precision  experiments, such as those involving optical lattice clocks \cite{Safronova:18}.

Quantum coherent control is a well-established  approach  free of these limitations, whereby quantum interference of transition pathways from an initially prepared coherent superposition of molecular states  is used to maximize or minimize the transition amplitudes, as in the classic Young double-slit experiment \cite{Book_Brumer,Scholak2017}. While coherent control has enjoyed great success when applied to unimolecular processes (such as photodissociation), its application to bimolecular collision dynamics has been limited by the need to entangle the internal and external degrees of freedom of collision partners, a significant experimental challenge \cite{Shapiro1996,Book_Brumer,Gong2003} that can be circumvented by using superpositions of {\it degenerate} magnetic sublevels ($m$-superpositions) as initial scattering states \cite{Brumer1999,Omiste2018,Devolder2020}. 
However, the  extent of control observed in previous studies of Penning and associative ionization in cold Ar~+~Ne$^*$ collisions \cite{Omiste2018}, atom-atom collisions in a magnetic field \cite{Herrera2008}, and differential F~+~H$_2$  reactive scattering \cite{Devolder2020} has been limited  due to large uncontrollable incoherent  terms or to symmetry reasons. 
As the general conditions for efficient coherent control of ultracold collisions are unexplored, it remains  unclear whether quantum interference-based control methods can be used to tune ultracold collision dynamics with an efficiency exceeding  that of their traditional dc  counterparts.

Here, we recognize quantum coherent control as an important  approach to manipulating ultracold molecular collisions. We show that  by forming coherent superpositions of initial molecular states, it is possible to achieve complete control over integral scattering cross sections and branching ratios in the $s$-wave regime of both the initial and final collision channels (the double $s$-wave regime).  Using rigorous quantum scattering calculations \cite{Tscherbul2008} we demonstrate  extensive  control of ultracold O$_2$~+~O$_2$ collisions, a system recently observed experimentally in a magnetic trap \cite{Segev:19}, over the unprecedented range of ten orders of magnitude.
   Our coherent control scenario does not require external electromagnetic fields   and can be applied to a wide range of atomic and molecular collisions that are not amenable to external dc field control,  such as those involving H$_2$ \cite{Gong2003}
   and homonuclear alkali-metal dimers.  This significantly expands the toolbox of methods for manipulating ultracold molecular collisions.




\color{black}

{\it Theory.} 
As a first step to achieving coherent control of cold collisions, we prepare an initial coherent superposition of $N_s$ two-molecule internal states $\ket{a_i b_i}$, where $a_i$ and $b_i$ denote internal states of each of the two colliding molecules:
\begin{equation}
\ket{\psi_{s}}=\sum_{i=1}^{N_{s}} c_i \ket{a_ib_i}.
\label{sup}
\end{equation}
Using the standard expression for the state-to-state integral cross section (ICS) $\sigma_{ab \rightarrow a'b'}=\frac{\pi}{k^2}\sum_{\ell,m_\ell}\sum_{\ell',m_\ell'} |T_{ab\ell,m_\ell \rightarrow a'b'\ell'm'_\ell}|^2$, where $\ell$ and $\ell'$ are the initial and final orbital angular momenta  of the collision,  $m_\ell$ and $m'_\ell$ are the projections of $\ell$ and $\ell'$ on the space-fixed quantization axis $Z$, $k$ is the initial relative momentum, and $ T_{ab\ell m_\ell \rightarrow a'b'\ell'm'_\ell}$ are the $T$-matrix elements, we obtain 
 the cross-section for scattering from the initial superposition (\ref{sup}) to the final two-molecule internal state $\ket{a'b'}$ as
\begin{equation}
\sigma_{s \rightarrow a'b'}=\frac{\pi}{k^2}\sum_{\ell,m_\ell}\sum_{\ell',m'_\ell} \left|\sum_{i=1}^{N_{s}} c_i T_{a_i b_i\ell m_\ell \rightarrow a'b'\ell'm'_\ell}\right|^2.
\label{cross_section_sup}
\end{equation}
For a  superposition of two states ($N_s$=2), this expression is analogous to that describing a set of double slit experiments for each partial wave pair $\{\ell m_\ell,\ell'm_\ell'\}$, where the interfering paths are defined by the initial collision channels, $\ket{a_1 b_1\ell m_\ell}$ and $\ket{a_2 b_2 \ell m_\ell}$ leading to the final channel $\ket{a'b'\ell'm'_\ell}$. The amplitudes of each path are $|c_1 T_{a_1b_1\ell,m_\ell \rightarrow a'b'\ell'm'_\ell}|$ and  $|c_2 T_{a_2 b_2 \ell m_\ell \rightarrow a'b'\ell' m'_\ell}|$, {and the relative phase between the paths is given by the sum of the relative phase of superposition coefficients $\beta$ (see below)  and that of the $T$-matrix elements ($\delta_{1}^{\{\ell m_\ell,\ell'm_\ell'\}}-\delta_{2}^{\{\ell m_\ell,\ell'm_\ell'\}}$)}. Because there is no interference between the terms with different $\ell$, $m_\ell$, $\ell'$, and $m_\ell'$ in Eq.~(\ref{cross_section_sup}), the efficiency of coherent control of the ICS depends on how well we can control the individual partial wave contributions. Thus, we expect the control efficiency to be strongly enhanced at low temperatures, when only a limited  number of  initial and final partial wave terms  are present in Eq. (\ref{cross_section_sup}).  In particular,  in the limit of zero collision energy, only $s$-wave terms with $\ell=0,m_{\ell}=0$   contribute to the ICS due to the Wigner threshold law \cite{Sadeghpour:00,Krems:05}, and the number of partial waves in the final channel is often strongly limited by angular momentum conservation \cite{Volpi:02,Krems:04}. 

Below we consider the case of nearly thermoneutral spin-exchange
 collisions dominated by $s$-waves in both the incident and final scattering channels ($\ell=\ell'=0)$, where the above arguments lead us to expect a large extent of coherent control.

Consider then a coherent superposition of two incident $s$-wave channels $\ket{a_1b_1 00}$  
and $\ket{a_2 b_200}$, which allows for coherent control of the ICS to the final $s$-wave channel $\ket{a'b'00}$. {Equation~(\ref{cross_section_sup})  then reduces to
\begin{equation}
\sigma_{s \rightarrow a'b'}=\frac{\pi}{k^2} \left| \cos \eta  T_{1} + \sin \eta e^{i\beta}  T_{2} \right|^2,
\label{cross_section_s_s}
\end{equation}
where we define $c_1=\cos \eta$, $c_2=\sin \eta e^{i\beta} $, $T_{1}=T_{a_1b_1 00 \rightarrow a'b'00}$ and $T_{2}=T_{a_2 b_2 00 \rightarrow a'b'00}$. Note that $\eta$ defines the relative population of each state in the superposition while $\beta$ gives the relative phase between the states.} 

The values of $c_1$ and $c_2$ that extremize the ICS can be found by  diagonalizing  the matrix $\mathcal{T}_{ij}=T_{i}T^*_{j}$ \cite{Frishman1999a}. 
The lowest eigenvalue corresponds to $\sigma_{s \rightarrow a'b'}^\text{min}=0$, which shows that {\it it is  always possible to coherently suppress collision-induced transitions to any given final channel $\ket{a'b'00}$} regardless of the values of $T_1$ and $T_2$.
The optimal values of the superposition parameters $\eta$ and $\beta$ that minimize the ICS are given by 
\begin{align}
\eta_\text{min}&= \cos^{-1}\left[\sqrt{{\sigma_{2}}/(\sigma_{1}+\sigma_{2}})\right] =\tan^{-1}(\sqrt{{\sigma_{1}/\sigma_{2}}} )\label{eta_min}  \\
\beta_\text{min}&=(\delta_{2}-\delta_{1})-\pi,
\label{beta_min}
\end{align}
where  $\sigma_{1}=\frac{\pi}{k^2}|T_1|^2$ and $\sigma_{2}=\frac{\pi}{k^2}|T_2|^2$ are the $s$-wave ICS for the incident  channels $\ket{a_1b_100}$ and $\ket{a_2b_200}$.

We observe that the optimal value of $\eta$ depends on the ratio of the s-wave ICS while the relative phase $\beta$ depends on the difference in phases of the $T$-matrix elements.  These results
can be understood using the analogy with the double slit experiment: The angle $\eta$ determines the distinguishability between the two paths while the angle $\beta$ controls their relative phase. The paths interfere destructively when they are indistinguishable and when the relative phase between them is a multiple of $\pi$. 
Indeed, at $\eta=\eta_\text{min}$, the two paths are indistinguishable with equal amplitudes, i.e., $\cos(\eta_\text{min})|T_1|=\sin(\eta_\text{min})|T_2|$, and the relative phase between the two paths $\beta+(\delta_{1}-\delta_{2})$ reaches $\pi$ when $\beta=\beta_\text{min}$ [Eq. (\ref{beta_min})].

From the second eigenvalue of $\mathcal{T}_{ij}$, we obtain the maximum value of the ICS, which is given by the sum of the ICSs from the initial channels  $\ket{a_1b_100}$ and  $\ket{a_2b_200}$ 
\begin{equation}
\sigma_{s \rightarrow a'b'}^\text{max}=\sigma_{1}+\sigma_{2}.
\end{equation}
Using coherent control, it is therefore possible to tune the $s$-wave ICSs between zero and $\sigma_{1}+\sigma_{2}$. As  $\sigma_1$ and $\sigma_2$ can reach very large values near collision thresholds  \cite{Sadeghpour:00,Krems:05,Balakrishnan:16}
a very wide control range is possible, as shown below for O$_2$~+~O$_2$ collisions. The superposition angles $\eta$ and $\beta$ that  maximize the ICS are given by 
\begin{align}  \label{eta_max}          
\eta_\text{max}&=\cos^{-1}\left[\sqrt{{\sigma_{1}}/(\sigma_{1}+\sigma_{2}})\right]=\tan^{-1}(\sqrt{{\sigma_{2}/\sigma_{1}}} ) \\
\beta_\text{max}&=\delta_{2}-\delta_{1}.
\label{beta_max}
\end{align}
Using the double-slit experiment analogy developed above,  the maximum ICS is obtained when   the relative phase between the two paths $\beta_\text{max}+(\delta_{1}-\delta_{2})$  is zero, leading to Eq.~(\ref{beta_max}).
In contrast to the case of $\sigma_{s \to a'b'}^\text{min}$, the best strategy for maximizing the ICS is to increase the distinguishability of the two interfering paths.
 Interestingly, the values of the superposition parameters that minimize and maximize the ICS are related by
$\eta_\text{max}+\eta_\text{min}=\pi/2$ and 
$\beta_\text{max}-\beta_\text{min}=\pi$.
\label{eq:rel_beta}

Having demonstrated complete coherent control of the total ICS, we now show that such control can  be extended to include the branching ratios ${\sigma_{s \rightarrow 1'}}/{\sigma_{s \rightarrow 2'}}$ for transitions to the final channels $|1'\rangle=\ket{a'_1b'_100}$ and $|2'\rangle=\ket{a'_2b'_200}$.
As shown above, there exists a superposition, defined by the parameters $\eta_\text{min}^{1'}$ and $\beta_\text{min}^{1'}$, for which the $s$-wave ICS  $\sigma_{s \rightarrow 1'}$ vanishes. Similarly, there is a superposition with the parameters $\eta_\text{min}^{2'}$ and $\beta_\text{min}^{2'}$, for which the ICS  $\sigma_{s \rightarrow 2'}$ vanishes. Then, the ICS ratio ${\sigma_{s \rightarrow 1'}}/{\sigma_{s \rightarrow 2'}}$  can be varied from zero to  infinity by tuning the superposition parameters from ($\eta_\text{min}^{1'}$, $\beta_\text{min}^{1'}$) to ($\eta_\text{min}^{2'}$, $\beta_\text{min}^{2'}$), thus achieving complete control over the branching ratio.


{\it Application: Coherent control of ultracold molecular collisions.} 
As an example consider the coherent control of ultracold collisions of O$_2 (X^3\Sigma)$ molecules in their ground electronic and rovibrational states ($v=N=0$, where $v$ is the vibrational quantum number and $N$ is the quantum number related to the square of the rotational angular momentum $\hat{N}^2$).
  Cold and ultracold O$_2 (X^3\Sigma)$ ~+~O$_2 (X^3\Sigma)$ collisions were studied theoretically by several groups  \cite{RerezRios:10,Tscherbul2008,Avdeenkov:01,Lepers:20} and have recently observed experimentally in a magnetically trapped oxygen gas  \cite{Segev:19}.
Due to their nonzero electron spin $S=1$, O$_2 (X^3\Sigma)$ molecules
can occupy three different spin states $\ket{M_S}$ with $M_S=-1,0,$ and 1 (assuming $S=1$ and neglecting the hyperfine structure for simplicity). {An inelastic collision can change  the spin projection of  one or both molecules i.e.,  $\ket{{M_A,M_B}}_p\rightarrow \ket{{M'_A,M'_B}}_p$,
where 
\begin{equation}
\ket{{M_A,M_B}}_p=\frac{1}{\sqrt{2(1+\delta_{M_AM_B})}}\left[\ket{M_A,M_B}+p \ket{M_B,M_A} \right].
\end{equation}
These are internal states of the colliding molecules that have been identical particle symmetrized and that include the parity $p$ of the state\cite{Tscherbul2008}. When $M_A=M_B$, only the state for $p=1$ exists while the both parities are possible otherwise. In this paper, we drop the index $p$, writing $\ket{{M_A,M_B}}$, when the calculated quantity (ICS or branching ratio) includes a sum on partial waves and then on the both parities.}

More specifically, consider the nearly thermoneutral spin-exchange collisions {$\ket{{0,0}}_p \leftrightarrow \ket{{-1,+1}}_p$}, which can be used to generate entanglement \cite{Duan:02} and quantum many-body phases in spinor Bose-Einstein condensates \cite{Chang:05,Kawaguchi2012,StamperKurn:13} and play an important role in ultracold atom-molecule and atom-ion chemistry \cite{Rui:17,Liu:19,Sikorsky:18}. At ultracold temperature, these flip-flop collisions  occur in  the $s$-wave regime for both the incident and final channels,  thus forming an ideal testing ground for the application of the coherent control theory developed above.  


 To calculate the $T$-matrix elements for ultracold O$_2$~+~O$_2$ collisions, we employ a rigorous time-independent quantum scattering approach \cite{Tscherbul2008} using an uncoupled symmetrized space-fixed  basis set composed of  three rotational states ($N=0{-}4$) and 6 partial waves ($\ell=0{-}10$).
 The coupled-channel equations were integrated on the radial grid  from 4.0 to 150 $a_0$, with a step of 0.04 $a_0$ to obtain  the $T$-matrix  elements converged to within $5\%$. {The calculation includes both parities $p=\pm1$}.
 The calculated $T$-matrix elements are then used to construct the ICSs [Eq. (\ref{cross_section_sup})]. The sum on partial waves is carried out in order to study collisions outside the double $s$-wave regime. 
 


To coherently control the spin-exchange ICS to the final channels {$\ket{{0,0}}$  and $\ket{{-1,+1}}$}, consider three different kinds of coherent superpositions of the initial molecular  spin states {$\ket{{0,0}}_p$} and {$\ket{{-1,+1}}_p$}. In particular, 
{\it an entangled two-molecule superposition}
\begin{equation}\label{psiE}
\ket{\psi_{E}}=\cos\eta \ket{{-1,+1}}_{p=\pm1}+\sin\eta e^{i\beta}\ket{{0,0}}_{p=+1}
\end{equation}
cannot be represented as a direct product of the individual molecules's states. While this superposition is the simplest to consider from a theoretical perspective, and  provides robust control (see below), it is challenging to prepare experimentally, as it requires entangling the internal states of the colliding molecules. Note that only the positive partity ($p=1$) gives a superposition of the two states. For a negative parity, $\ket{\psi_{E}}$ reduces to $\ket{{-1,+1}}_{-1}$ and gives uncontrollable odd-partial wave contributions to the ICS. 

{\it A non-entangled initial superposition} has the form of a tensor product of two  single-molecule superposition   states $\ket{\psi_A}\ket{\psi_B}$, where
\begin{align}
\ket{\psi_A}&=N_2\left(\sqrt{\cos\eta} \ket{-1}+\sqrt{\sin\eta}e^{i\frac{\beta}{2}}\ket{0}\right) \\
\ket{\psi_B}&=N_2\left(\sqrt{\sin\eta}e^{i\frac{\beta}{2}}\ket{0}+\sqrt{\cos\eta} \ket{+1}\right),
\end{align}
where $N_2=(\sin\eta+\cos\eta)^{-1/2}$.  For two identical bosonic molecules such as O$_2$, this initial state must be symmetrized to account for identical particle permutation symmetry \cite{Tscherbul2008} giving {
	\begin{multline}\label{psiS2}
	\ket{\psi_{2}^S}=N_2^2 \Big[\cos\eta \ket{{-1,+1}}_{p=\pm1}+\sin\eta e^{i\beta}\ket{{0,0}}_{p=+1}\\+\sqrt{\cos\eta\sin\eta}e^{i\frac{\beta}{2}} (\ket{{-1,0}}_{p=\pm1}+\ket{{0,+1}}_{p=\pm1})\Big].
	\end{multline}}
	This initial state can be created in, e.g., merged beam experiments \cite{NareviciusScience12} by preparing coherent superpositions of  internal states of the individual molecules prior to collision.
	In a similar way, we can prepare  a {\it non-entangled three-state superposition}
	\begin{equation}\label{psiS3}
\ket{\psi_{A}}=N_3 \left[\sqrt{\cos\eta}(\ket{-1}+\ket{+1})+\sqrt{\sin\eta}e^{i\frac{\beta}{2}} \ket{0}\right],
\end{equation}
where $N_3=({\sin\eta+2\cos\eta})^{-1/2}$. 
After symmetrization, the initial wavefunction becomes
{ 
\begin{equation}
\begin{split}
\ket{\psi_{3}^S}=N_3^2\Big[&\cos\eta \ket{{-1,+1}}_{p=\pm1}+\sin\eta e^{i\beta}\ket{{0,0}}_{p=+1}\\+&\sqrt{\cos\eta\sin\eta}e^{i\frac{\beta}{2}} (\ket{{-1,0}}_{p=\pm1}+\ket{{0,+1}}_{p=\pm1})\\ +& \cos\eta (\ket{{-1,-1}}_{p=+1}+\ket{{+1,+1}}_{p=+1}) \Big].
\end{split}
\end{equation}
A key difference between the entangled and non-entangled superpositions is
 the presence of the uncontrolled ``{\it satellite terms}'' [see Ref. \cite{Book_Brumer}] $\ket{{-1,0}}_{p=\pm1}$ and $\ket{{0,+1}}_{p=\pm1}$ in $\ket{\psi_{2}^S}$ and $\ket{{-1,0}}_{p=\pm1}$, $\ket{{0,+1}}_{p=\pm1}$, $\ket{{-1,-1}}_{p=+1}$, and $\ket{{+1,+1}}_{p=+1}$ in $\ket{\psi_{3}^S}$.}

	\begin{figure}
	\centering
	\includegraphics[width=0.9\columnwidth, trim = 40 20 0 170]{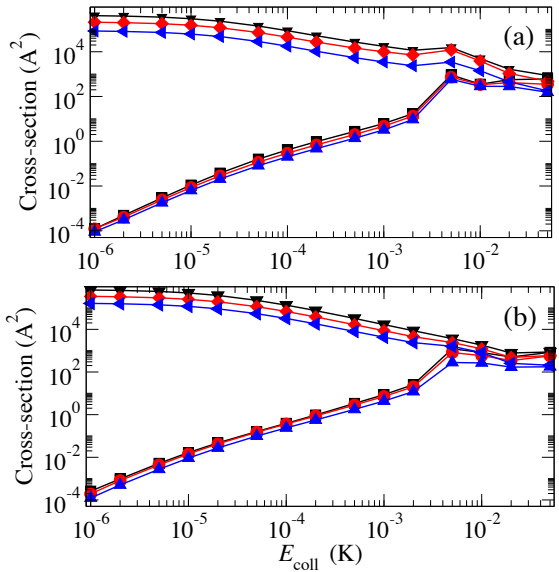}
	\caption{Minimum (lower traces) and maximum (upper traces) ICSs from the initial superpositions $\ket{\psi_{E}}$ (black), $\ket{\psi_{2}^S}$ (red) and $\ket{\psi_{3}^S}$ (blue) to the final collisional channel (a) {$\ket{{0,0}}$} and (b){$\ket{{-1,+1}}$}. }
	\label{fig_control_cross_section}
\end{figure}

Figure~\ref{fig_control_cross_section} shows the minimum and the maximum values of  the ICS obtained with the initial superpositions $\ket{\psi_E}$, $\ket{\psi_{2}^S}$ and $\ket{\psi_{3}^S}$ to the final channels {$\ket{{0,0}}$} and {$\ket{{-1,+1}}$} as a function of collision energy $E_\text{coll}$. The values for $\eta$ and $\beta$ were determined by the  Eqs. (\ref{eta_min}), (\ref{beta_min}), (\ref{eta_max}), and (\ref{beta_max}). A remarkably wide, nine orders of magnitude range of control is observed for both final states. We further observe from Fig.~\ref{fig_control_cross_section} that the vast extent of coherent control in the $s$-wave regime is insensitive to whether the initial superposition is chosen to be entangled or non-entangled.  
The ICSs obtained for the entangled superposition (\ref{psiE}) are slightly larger than those for the non-entangled superpositions (\ref{psiS2}) and  (\ref{psiS3}) due to the different normalization of the controllable term {$\cos\eta \ket{{-1,+1}}_{p=\pm1}+\sin\eta e^{i\beta}\ket{{0,0}}_{p=+1}$} in these two superpositions.
The $s$-wave to $s$-wave contribution due to this term is suppressed and the value of the ICS is determined by $\ell \ge 2$ partial waves of the controllable term, as well as by the satellite terms.
 The increase of $\sigma^\text{min}$ with increasing collision energy  observed in Fig.~\ref{fig_control_cross_section}  is due to  the growing contributions  of partial waves with $\ell\ge 2$ and of the satellite terms to the ICS, which, as noted above, diminish the extent of control. At $E_\text{coll}<$~5~mK, the $s$-wave to $s$-wave contribution to the ICS exceeds 99~\% making the total ICS fully controllable.  In contrast, at collision energies above the height of the $\ell=2$ centrifugal barrier ($E_\text{coll} > 5$~mK) the non-$s$-wave contributions become dominant and the degree of coherent control is sharply diminished. Also, at $E_\text{coll} > 5$~mK, the satellite terms reach substantial values, as these terms are related to spin exchange processes, which change the value of the total angular momentum projection $M=M_A+M_B$, and thus require $\ell\ge 2$ to occur. Note that the non-$s$-wave and satellite contributions  go to zero at 0 K, in which case the minimum cross-section $\sigma^\text{min}$ is exactly zero. 



\begin{figure}
	\centering
	\includegraphics[width=0.7\columnwidth, trim = 20 20 0 20]{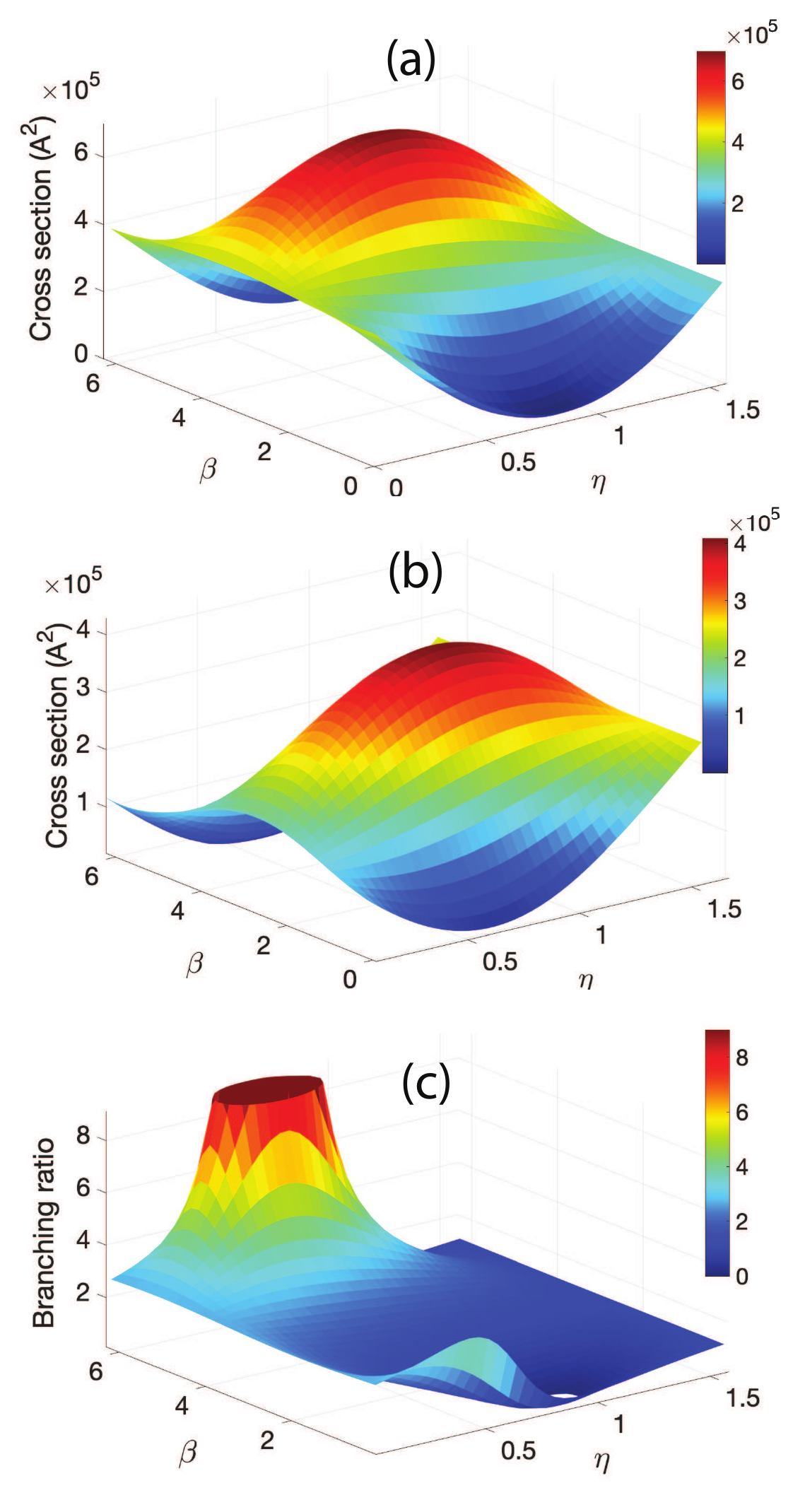}
	\caption{Coherent control of the ICSs $\sigma_{s\to a'b'}$ for ultracold O$_2$~+~O$_2$ collisions starting from the initial superposition $\ket{\psi_{2}^S}$ as a function of the superposition parameters $\eta$ and $\beta$ at a collision energy of 1~$\mu$K in the absence of external fields. The final states are {$\ket{{-1,+1}}$} (a)  and {$\ket{{0,0}}$} (b). Panel (c) shows the branching ratio  ${\sigma_{s \rightarrow -1+1 }}/{\sigma_{s \rightarrow 00}}$. While the values shown are limited to 8 to aid visibility,  the maximal value of the branching ratio is $1.2\times 10^{9}$.}
	\label{fig_xs_etabeta}
\end{figure}

Figure \ref{fig_xs_etabeta} (a) and (b) shows the ICS for ultracold  O$_2$~+~O$_2$ collisions as a function of the initial superposition parameters  $\eta$ and $\beta$. In addition to the wide range of control for both the final spin exchange channels {$\ket{{-1,+1}}$} and{$\ket{{0,0}}$}, we note that it is possible to tune the ICS in a continuous manner, reaching all intermediate values between zero and $\sigma_\text{max}$ by varying the superposition angles $\eta$ and $\beta$. The dependence of the ICS on $\eta$ and $\beta$ exhibits characteristic oscillations given by Eq.~(\ref{cross_section_s_s}), which can be recognized as a signature of coherent control.

\begin{figure}
	\centering
	\includegraphics[width=0.96\columnwidth, trim = 20 20 0 60]{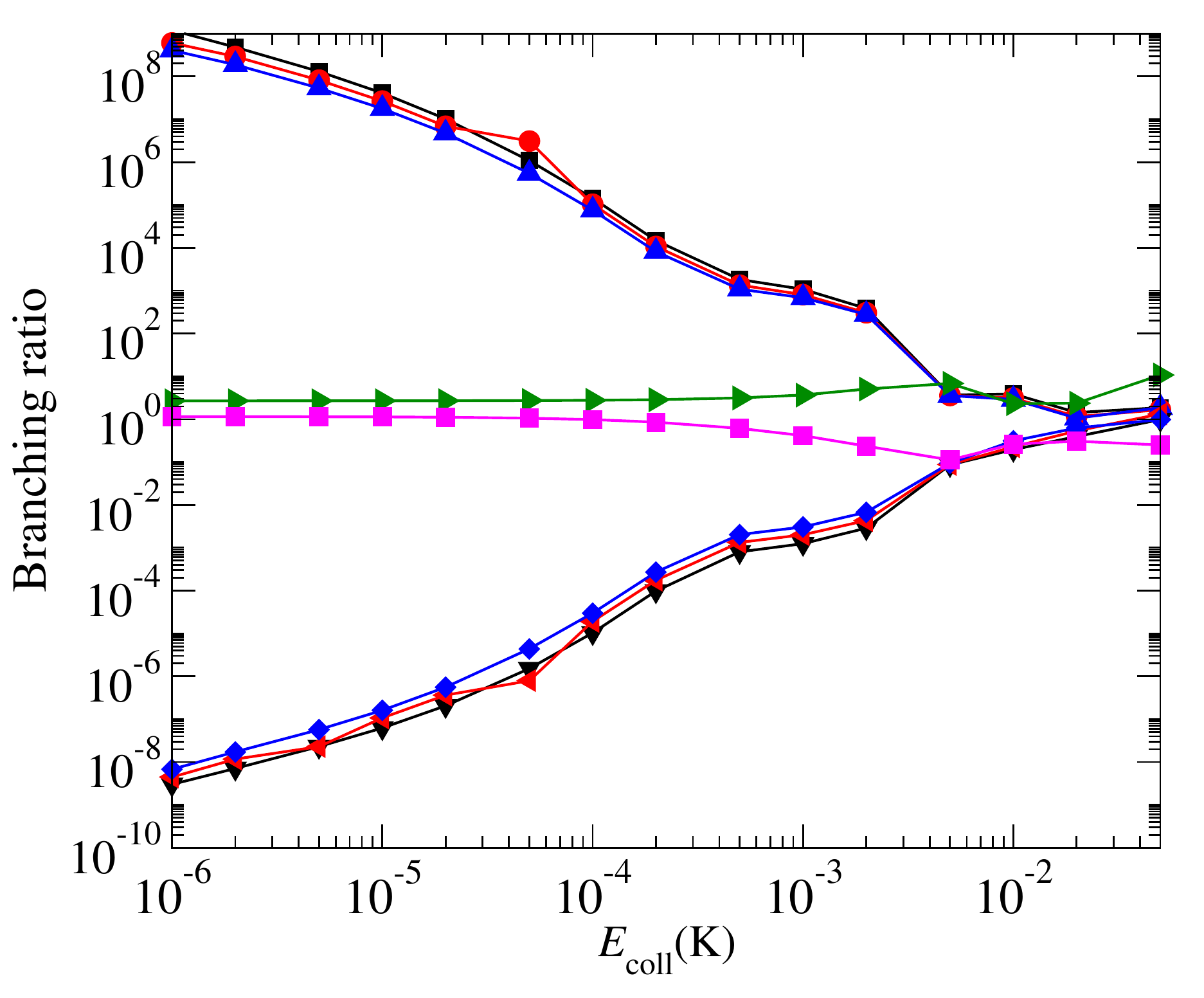}
	\caption{Minimum (lower traces) and maximum (upper traces) of the branching ratio  ${\sigma_{s \rightarrow -1+1 }}/{\sigma_{s \rightarrow 00}}$ for the initial superpositions $\ket{\psi_{E}}$ (black), $\ket{\psi_{2}^S}$ (red) and $\ket{\psi_{3}^S}$ (blue). The branching ratios in the absence of control are also shown as middle traces for the initial states {$\ket{{-1,+1}}$} (triangles) and {$\ket{{0,0}}$} (squares).}
	\label{fig_branch_ratio}
\end{figure}

Finally, consider coherent control of the branching ratio ${\sigma_{s \rightarrow -1+1}}/{\sigma_{s \rightarrow 00}}$, which is minimized when the ICS $\sigma_{s \rightarrow -1+1}$ is minimized and maximized when  $\sigma_{s \rightarrow 00}$ is minimized. At 1 $\mu$K, for example, the branching ratio can be varied from $10^{-9}$ to $10^8$ demonstrating a truly outstanding range of control, spanning seventeen orders of magnitude! As a reference, the branching ratios in the absence of control are 2.69 and 1.15 for the initial states {$\ket{{-1,+1}}$} and {$\ket{{0,0}}$}. Figure \ref{fig_xs_etabeta} (c) shows the branching ratio as a function of the initial superposition parameter. A sharp peak around the maximal value is observed. 
 The range of control observed in Fig.~\ref{fig_branch_ratio} is much wider than in any previous study of coherent control \cite{Book_Brumer,Omiste2018,Devolder2020}, showing that the ultracold $s$-wave threshold regime provides optimal conditions for coherent control of quantum scattering dynamics. As in the case of the ICS, we observe a gradual loss of control as the collision energy is increased until control is completely lost outside of the $s$-wave  regime at $E_\text{coll}>5$~mK.

In conclusion, we have developed a general theory of quantum interference-based coherent control of ultracold collisions, which allowed us to establish the possibility of complete coherent control over quantum scattering in the regime where only a single partial wave  is involved in both the incident and final collision channels. We show that ultralow temperatures strongly enhance coherent control by favoring $s$-wave threshold scattering,  
and we determine the optimal parameters of the coherent superpositions required to maximize and minimize  the ICS.
The theory was applied to control ultracold spin-exchange collisions of oxygen molecules. We demonstrate vast control over both the  ICS and their branching ratios  in the $s$-wave threshold regime.
Our results demonstrate the  possibility of using quantum interference as a powerful tool for controlling ultracold collision dynamics, which can be applied to a much wider range of molecular species (such as H$_2$) than dc field control.  
We envision examining the possible use of coherent control to tune spin-exchange interactions in spinor Bose-Einstein condensates  \cite{Chang:05,Duan:02,StamperKurn:13} and to control cold atom-molecule and atom-ion spin-exchange chemistry \cite{Rui:17,Liu:19,Sikorsky:18}. 
It would also be interesting to explore the possibility of coherently controlling exothermic relaxation processes in ultracold collisions outside the $s$-wave threshold regime for the final  collision channel, especially when a single partial wave is implied due to the presence of a resonance \cite{Jambrina:19,Morita:20}.

This work was supported by the U.S. Air Force Office for Scientific Research (AFOSR) under Contract No.
FA9550-19-1-0312. SciNet computational facilities are gratefully acknowledged.

 
\bibliography{Ultracold_coherent_control}
\end{document}